\documentclass[12pt,preprint]{aastex}

\usepackage{graphicx,epsfig,natbib,color}
\usepackage{lscape}
\usepackage{graphicx}
\usepackage{epsfig}
\usepackage{natbib}
\usepackage{multirow}

\begin{document}
\title{On the Origin of the EUV Late Phase of Solar Flares}

\author{Kai Liu\altaffilmark{1,2}, Jie Zhang\altaffilmark{2}, Yuming Wang\altaffilmark{1}, Xin Cheng\altaffilmark{2,3}}

\affil{$^1$ School of Earth and Space Science, University of Technology and Science of China, Hefei 230026, China}\email{kailiu@mail.ustc.edu.cn}
\affil{$^2$ School of Physics, Astronomy and Computational Sciences, George Mason University, 4400 University Drive, MSN 6A2, Fairfax, VA 22030, USA}
\affil{$^3$ Department of Astronomy, Nanjing University, Nanjing 210093, China}
\affil{\large{The Astrophysical Journal; Accepted: March 27, 2013}}

\begin{abstract}

Solar flares typically have an impulsive phase that followed by a gradual phase as best seen in soft X-ray emissions. A recent discovery based on the EUV Variability Experiment (EVE) observations onboard the Solar Dynamics Observatory (\emph{SDO}) reveals that some flares exhibit a second large peak separated from the first main phase peak by tens of minutes to hours, which is coined as the flare's EUV late phase \citep{Woods2011}. In this paper, we address the origin of the EUV late phase by analyzing in detail two late phase flares, an M2.9 flare on 2010 October 16 and an M1.4 flare on 2011 February 18, using multi-passband imaging observations from the Atmospheric Imaing Assembly (AIA) onboard \emph{SDO}. We find that: (1) the late phase emission originates from a different magnetic loop system, which is much larger and higher than the main phase loop system. (2) The two loop systems have different thermal evolution. While the late phase loop arcade reaches its peak brightness progressively at a later time spanning for more than one hour from high to low temperatures, the main phase loop arcade reaches its peak brightness at almost the same time (within several minutes) in all temperatures. (3) Nevertheless, the two loop systems seem to be connected magnetically, forming an asymmetric magnetic quadruple configuration. (4) Further, the footpoint brightenings in UV wavelengths show a systematic delay of about one minute from the main flare region to the remote footpoint of the late phase arcade system. We argue that the EUV late phase is the result of a long-lasting cooling process in the larger magnetic arcade system.

\end{abstract}


\section{Introduction}
Solar flares, one of the most energetic phenomena in the solar system, have remained a subject of intense study ever since they were first discovered more than one and half centuries ago by \citet{Carrington1859}. A solar flare is often observed as a sudden and rapid enhancement in intensity in almost all electromagnetic wavelengths due to a massive magnetic energy release in the Sun's corona resulting in particle acceleration and plasma heating. The present standard model of solar flares, the so-called ``CSHKP" model \citep{Svestka1992}, based on ideas of \citealt{Carmichael1964,Sturrock1968,Hirayama1974} and \citealt{Kopp1976}, is centered around the physical mechanism of magnetic reconnection in a bipolar magnetic flux system.

According to the standard model, the fast magnetic reconnection in a current sheet (CS) releases the magnetic energy stored in the corona and produces the observed flare. The reconnecting current sheet is probably formed though the eruption of a magnetic flux rope, as inferred from the presence of filaments and sigmoids embedded at the core of the bipolar source region. The standard model explains well several commonly observed phenomena: in particular, the separation motion of flare ribbons in the chromosphere and the slow-rising motion of post-flare loop arcades in the corona \citep[e.g.][]{Priest2002}.

The temporal evolution of a solar flare emission usually has two well-defined phases visible in its X-ray signature: the impulsive phase and the gradual phase \citep[e.g.][]{Benz2008,Hudson2011}. The impulsive phase is characterized by the intense increase of electromagnetic emissions lasting from a few minutes to tens of minutes, in particular in non-thermal signatures such as hard X-rays; this impulse usually corresponds to the rise phase of the flare's soft X-ray profile \citep{Neupert1968,Kane1980}. Immediately following the impulsive phase the soft X-ray emission, dominated by thermal radiation, tends to gradually return to its original level in a smooth and monotonic way. This gradual phase is also called the decay phase of a flare \citep{Moore1980}.

Recently, \citet{Woods2011} discovered a new phase of solar flare evolution using EUV irradiance observations on the EUV Variability Experiment (EVE; \citealt{Woods2012}) on board the Solar Dynamics Observatory (\emph{SDO}; \citealt{Pesnell2012}). They found that, in addition to a peak during the gradual phase in EUV ``warm'' coronal emissions (e.g., $\sim$2.5 MK from Fe XVI 33.5nm), some flares also exhibit a second peak separated by tens of minutes to hours from the first peak; the second peak in EUV is completely absent in the usual smooth profiles of soft X-ray gradual phase (thus evading detection in the past). This second peak is coined as the EUV late phase by \citet{Woods2011}. They also pointed out that the late phase originates from a second set of higher loops within the flaring active region, i.e., where the fast magnetic reconnection occurs, can be called the flare main region, comparing to the flare late phase region. Most of the EUV coronal emissions from the main region always have a peak during the flare's gradual phase. Although different EUV emissions would have different peak times, for better description, we refer this first emission peak from the flare main region as the flare's main phase in generally.

In this paper, we intend to address the origin of the EUV late phase through studying two late phase flares that are well observed by the Atmospheric Imaging Assembly (AIA; \citet{Lemen2011}) also onboard \emph{SDO}. The high-cadence, high resolution and multi-temperature capacity of AIA enables us to observe the morphological structure of a flare and track the evolution of different parts of the structure. We find that the flare's main  phase originates from a very compact loop arcade, while the late phase from a separate and much larger loop arcade. Nevertheless, the two loop systems, each having two footpoints, have one footpoint sharing a common magnetic region, suggesting a morphological connection between the two magnetic loop systems. Further, the footpoint brightening of the late phase arcade in the remote region is tens of seconds lagging behind the footpoint brightenings in the compact flare region, indicating that there might be an energy sharing between the compact loops and the late phase loops. We also find that the time differences between the impulsive phase peak and the late phase peaks is a sensitive function of temperatures, indicating that the evolution of the late phase is dominated by a long-lasting cooling process. Observations and results are presented in Section 2, Discussions are given in Section 3, and Section 4 is for Summary and Conclusion.

\section{Observations and Results}

\subsection{Data}

The two flare events studied in this paper are selected from EVE observations, which measure the solar EUV differential irradiance from 0.1 nm to 105 nm with unprecedented spectral resolution of 0.1 nm, temporal cadence of 10 seconds and accuracy of 20$\%$ ~\citep{Woods2012}. The EVE Level-2 processing produces a combined set of merged spectra provided in a pair of files: one file contains the full spectra for one hour , and the other contains the isolated lines from the ionized solar elements and the bands simulated for other instruments such as AIA. In this paper, we use the EVE level-2 line data (EVL), which are publicly available from the website (http://lasp.colorado.edu/eve/data\_{}access/evewebdataproducts/level2/). The fits data files can be read by ``eve\_{}read\_{}whole\_{}fits.pro". The detailed morphological evolution of the flares are observed by the advanced AIA instrument, which takes high-resolution images ($4096\times4096$ pixels, 0.6" pixel size and 1.5" spatial resolution) of the whole Sun with a high cadence of 12 seconds. Further, the AIA instrument has ten passbands sensitive to different temperatures, thus allowing the investigation of the thermal property of flare structures. Six of the ten passbands are sensitive to coronal temperatures, which in the order of decreasing temperatures are 13.1 nm (Fe XX, $\sim$10 MK), 9.4 nm (Fe XII, $\sim$6.4 MK), 33.5 nm (Fe XVI, $\sim$2.5 MK), 21.1 nm (Fe XIV, $\sim$2.0 MK), 19.3 nm (Fe XII, $\sim$1.6 MK), and 17.1 nm (Fe IX, $\sim$0.7 MK), respectively. Note that the AIA 13.1 nm passband also contains a cool spectral line from Fe VIII, and the AIA 19.3 nm passband contains a hot spectral line from Fe XXIV. Nevertheless, such ``contamination" shall not affect the determination of the peak times of different flare phases, the main concern of this paper, since these phases well separates in time (thus in temperature). We also use the 160 nm passband (C IV, $\sim$0.1 MK) images, which observe the transition region and upper photosphere, to study the footpoint brightening of flares. The magnetic property of flare source regions is obtained from the Helioseismic and Magnetic Imager (HMI; \citealt{Scherrer2012}) also on board \emph{SDO}. We use standard software programs in the \emph{SolarSoftware} (\emph{SSW}) package to download AIA and HMI data. The "aia\_{}prep.pro" is used to calibrate the AIA fits file. Then the map structure technique (map package in \emph{SSW}) is applied to process the calibrated images, which includes plotting, cutting out, and de-rotating. We also use the same mapping method for the HMI data.

\subsection{M2.9 flare on 2010 October 16}

\begin{figure} 
     \vspace{-0.0\textwidth}    
     \centerline{\hspace*{0.02\textwidth}
               \includegraphics[width=1.0\textwidth,clip=1]{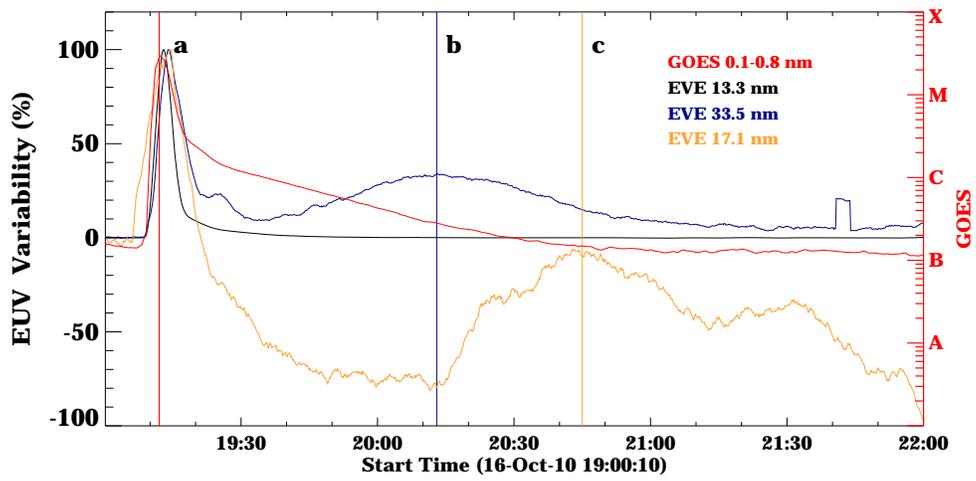}
               }

\vspace{0.0\textwidth}   
\caption{The temporal variation profiles of the irradiance in three different spectral lines of SDO EVE observation of the M2.9 flare on 2010 October 16. The GOES profile is also shown. Vertical lines a, b, and c indicate the flare's impulsive phase peak, EVE 33.5 nm late phase peak, and EVE 17.1 nm late phase peak, respectively. All profile in this paper are normalized as taking each one's pre-flare ($\sim$10 minutes before the beginning of impulsive phase) value to be 0 and its biggest peak value to be 1.}
\label{f1}

\end{figure}

The first event we studied is an M2.9 class flare that occurred on 2010 October 16. The M2.9 classification is based on GOES (Geosynchronous Operational Environment Satellite) measurement of the integrated X-ray flux from 0.1-0.8 nm. The X-ray intensity temporal profile of the flare (red line in Figure~\ref{f1}) shows that the rise phase (the rise phase in soft X-ray corresponds to the impulsive phase if non-thermal data are used) began at $\sim$19:07 UT and ended at $\sim$19:12UT (the peak time of soft X-rays is indicated by the vertical line a), lasting for only 5 minutes; the gradual phase or decay phase lasted for about 90 minutes before returning to the pre-flare level. Figure \ref{f1} also show the temporal profiles of the intensity of three EUV emission lines from EVE observations: 13.3 nm (Fe XX), 33.5 nm (Fe XVI) and 17.1 nm (Fe IX). While numerous lines are observed by EVE from 0.1 nm, these three lines are particularly chosen to represent the plasma temperatures of being hot ($\sim$10 MK), warm ($\sim$3 MK) and cool ($\sim$1 MK) respectively. Further, the three EVE lines have complementary imaging observations from AIA. The temporal profiles of the three EVE lines are similar to the soft X-ray profile only during the rise phase and also during the early decay phase. However, the light curves Fe XVI 33.5 nm and Fe IX 17.1 nm lines show a second intensity peak (the peak times are indicated by vertical lines b and c respectively), which is completely absent in the soft X-ray profile. The Fe XVI 33.5 nm profile shows a gradual-rise-gradual-fall second peak from $\sim$19:35 UT to $\sim$21:00 UT. The time of this second peak is 20:13 UT, which is about 60 minutes behind the impulsive phase peak in soft X-ray. Similar behavior is seen in Fe IX 17.1 nm, whose second peak time is delayed by almost 90 minutes from the impulsive phase peak.

\begin{figure} 
     \vspace{-0.0\textwidth}    
     \centerline{\hspace*{0.02\textwidth}
               \includegraphics[width=1.0\textwidth,clip=1]{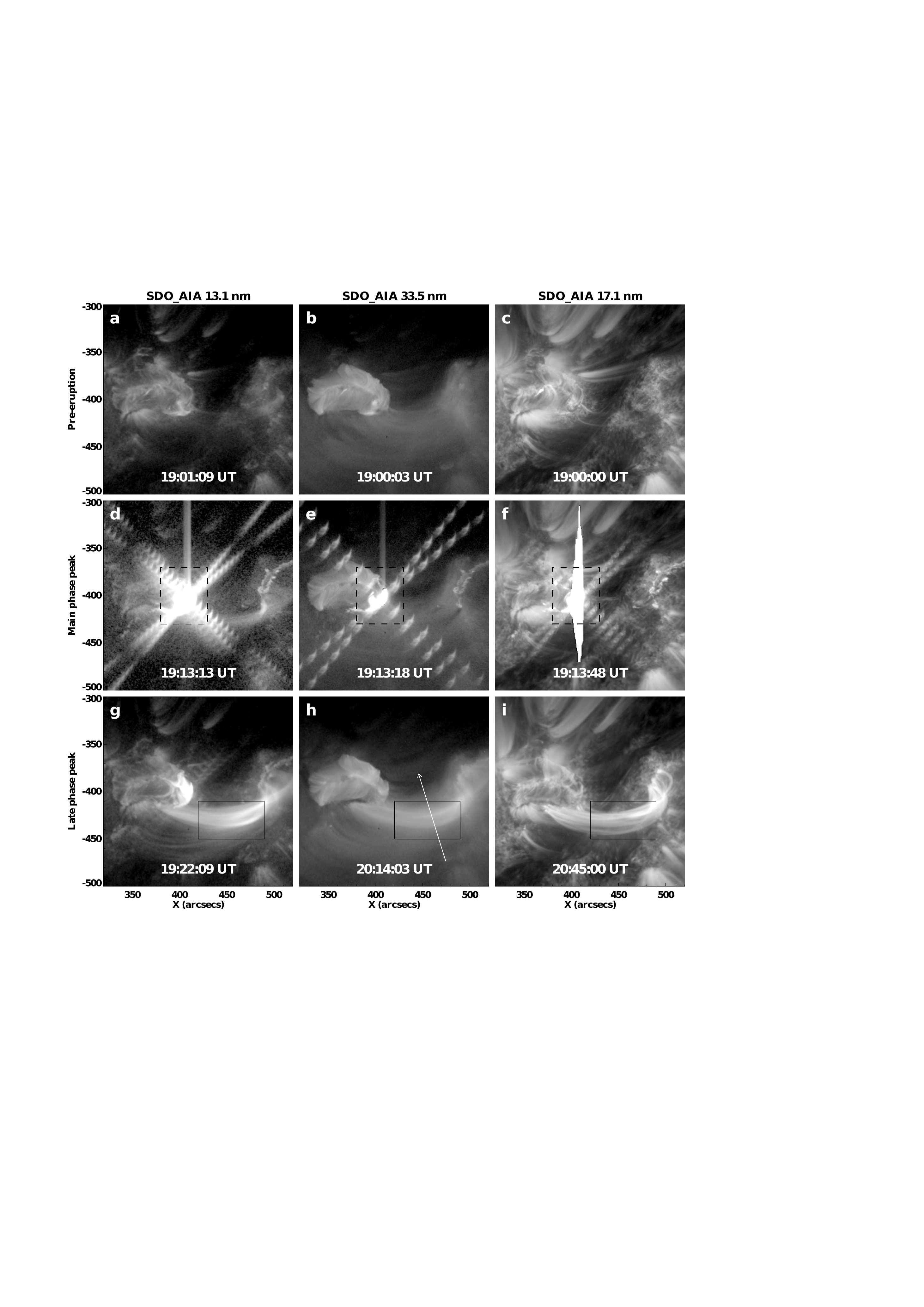}
                              }

     \vspace{0.0\textwidth}   
\caption{AIA sub-area images showing the structural evolution of the M2.9 flare on 2010 October 16. From left to right are AIA images in 13.1 nm, 33.5 nm and 17.1 nm passbands, respectively. (a)--(c): the pre-eruption images. (d)--(f): the images at the flare's main phase peak time. The black dotted-line boxes indicate the flare main region. (g)--(i): image at the peak times of the late phase (note that the times are different for different temperatures). The black solid-line boxes indicate the center region of the late phase arcade.}
   \label{f2}

\end{figure}

To determine the source region responsible for the late phase emission (as well as the main phase), we inspected AIA images covering the entire main phase and the late phase. The flare occurred in active region NOAA AR 11112, centered at the heliographic coordinates 20${^\circ}$S and 26$^{\circ}$W when the flare occurred. There was no other major activity on the disk during the period of this event, so the irradiance variations seen in EVE were largely produced by this single AR. For the following analysis, we have focused on a sub-area, small but containing the entire AR, of the full AIA images (as shown in Figure~\ref{f2}). At the peak time of the main phase, a small and compact area inside the AR (indicated by the black dotted-line box in Figure~\ref{f2}d, e, and f) exhibits an extreme brightening in all passbands of AIA images, indicating that this sub-region is the source of the flare's main phase.

\begin{figure} 
     \vspace{-0.0\textwidth}    
\centerline{\hspace*{0.0\textwidth}
               \includegraphics[width=1.0\textwidth,clip=1]{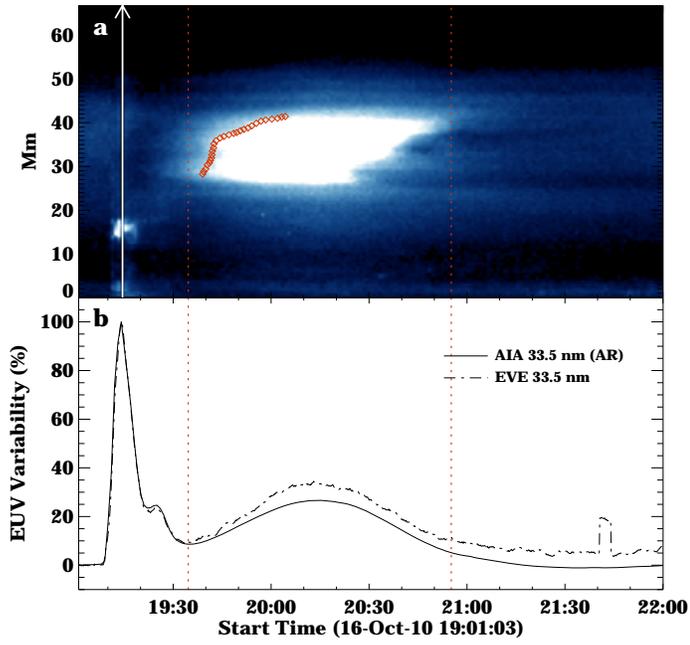}
            }

     \vspace{0.0\textwidth}   

\caption{(a): Slice-time plot of the AIA 33.5 nm images for the flare on 2010 October 16. The slice is indicated by the arrow line in Figure ~\ref{f2}h. (b): EUV variability from both EVE 33.5 nm and AIA 33.5 nm observations. The AIA flux is the total flux of the cut-out of the AR shown in Figure ~\ref{f2}}
   \label{f3}

   \end{figure}

We further inspect the sequence of AIA 33.5 nm images, looking for the source region of the late phase emission. We cut a slice of AIA 33.5 nm images (white arrow in Figure~\ref{f2}h) and plot its time evolution in Figure~\ref{f3}a. The light curves of EVE 33.5 nm emission and the AIA 33.5 nm emission summed over the cut-out of the AR are shown in Figure ~\ref{f3}b. Combing Figure~\ref{f3} and the inspection of the AIA image sequence, we find that a large loop arcade appeared at around 19:35 UT, lasting for about 90 minutes, then disappeared at about 21:01 UT (more accurately, cooling out of this AIA passband as discussed later); the central part of the loop arcade is indicated by the black box in Figure~\ref{f2}h. One can conclude that the time evolution of the emission from the loop arcade in AIA 33.5 nm images coincides very well with the temporal profile of the EVE spectral irradiance at 33.5 nm (as shown in Figure~\ref{f3}b). This temporal overlap proves straightforwardly that the large loop arcade is the source region of the EUV late phase brightening.

It is interesting to note that the late phase loop arcade is much different from the main phase loop arcade (which is called the gradual phase loop arcade in \citealt{Woods2011}'s paper), as shown in Figure~\ref{f2}. The late phase arcade consists of much longer loops; the span of the loops, as measured from the separation distance of the two footpoints, is about 120 arcsec, or 86,800 km in length on the projection. On the other hand, the main phase loop arcade is very compact. The separation of the two footpoints is about only 39 arcsec, or 28,200 km in length. Therefore, the linear size of the late phase arcade is about three times as large as that of the flare main emission arcade. Subsequently, the late phase loop arcade is much higher than the main phase loop arcade. The orientations of the two loop arcades are also different. While the late phase arcade is mainly east-west oriented, the main phase arcade is mainly oriented along the north-south direction. Further, the late phase arcade has showed apparent rising motion along the slice direction (displayed by the red diamonds in Figure~\ref{f3}a), which suggests that the late phase arcade participate in the magnetic reconnection. The rising motion in AIA 33.5 nm is particularly obvious during the rising time of the late phase, but appears stopped after the peak time of the late phase.

EVE data indicate that the late phase is most conspicuous in Fe XVI 33.5 nm emission. Nevertheless, the late phase also appears in other wavelengths. Figure~\ref{f2} panel g and i show that the late phase arcade also appears in AIA 13.1 nm and 17.1 nm images, respectively. The arcades in the hot and cool temperatures appear at the same location as in the warm temperature. However, the times of the appearances are much different. The large arcade first appeared in hot temperature in AIA 13.1 nm and then progressively appeared in cooler and cooler temperatures. The morphological evolution of the arcades with time in all six AIA corona passbands is shown in the online movie.

\begin{figure} 
     \vspace{-0.0\textwidth}    
\centerline{\hspace*{0.0\textwidth}
               \includegraphics[width=1.0\textwidth,clip=1]{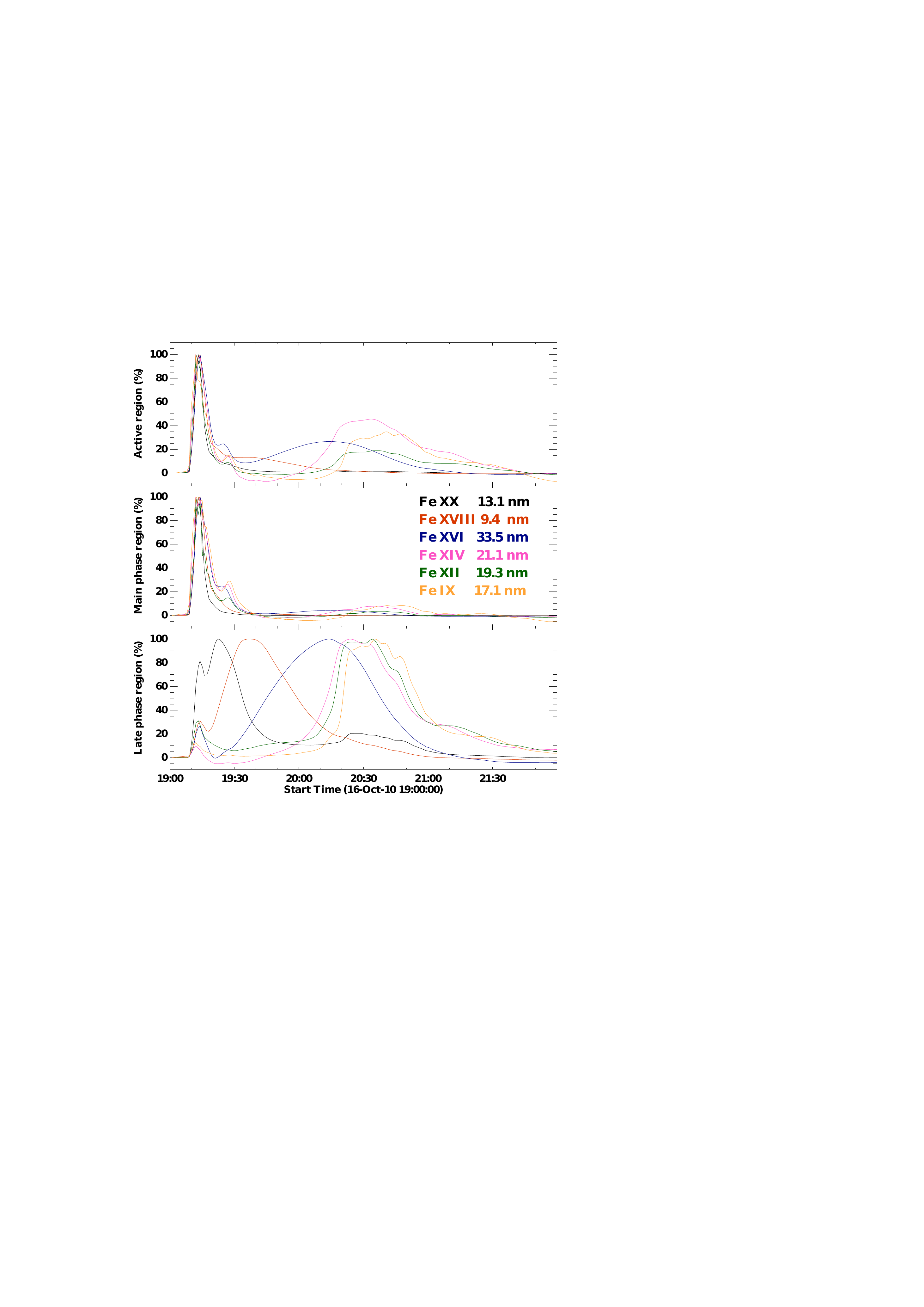}
            }

     \vspace{0.0\textwidth}   

\caption{Intensity temporal profiles of total EUV flux in selected regions of M2.9 flare on 2010 October 16 based on AIA images of six coronal passbands. The top panel shows the profiles of the whole active region. The middle and bottom panels show that from the flare main phase region and the flare late phase region, respectively.}
   \label{f4}

   \end{figure}

To further study the temporal evolution of the late phase arcade, as well as the main phase arcade, we make use of the high cadence AIA images to obtain the EUV flux variation of these specific features. In Figure~\ref{f4}, we show the variation of the total EUV flux of the entire active region (upper panel), the compact main phase arcade region (middle panel) and the large late phase arcade region (low panel), respectively. The sizes of the three regions used to sum up the total flux are indicated in Figure~\ref{f2}: the entire region, the region in the black dotted-line box, and the region in the black solid line-box, respectively. There are significant differences between the flux variation in the main phase arcade and the late phase arcade. First, the flux variation is much faster in the main phase than that in the late phase. The width of the main phase peak is of only several minutes, while that of the late phase peak lasts for tens of minutes long. Secondly, probably more importantly, the fluxes in all temperatures reach their peaks at almost the same time in the main phase arcade, but reach the peaks at quite different times in the late phase arcade.

To characterize the thermal property described above, we measure the peak times, the durations of the full-width-half-maximum (FWHM) of the flux profiles in all six coronal passbands for the two arcades; the values are listed in Table~\ref{t1} along with the ratio of peak intensities between the late phase and the main phase. Table~\ref{t1} shows that the main phase peak times for the six AIA EUV passbands are almost the same, considering that they are all within 13 seconds for a measurement uncertainty of 12 seconds (the cadence of AIA images). On the other hand, the peak times for the late phase, are much different. The time differences between the late phase peaks of the six passbands and the impulsive phase peak of the 13.1 nm passband (the highest temperature of AIA passbands), from high to low temperatures, are 9, 24, 60, 71, 81 and 92 minutes, respectively. The progressively short time delays in the hotter and hotter temperature passbands indicate that the onset times of the thermal emissions in the highest temperatures in the two magnetic systems should be very close, if not at the same time. Therefore, there must be a direct connection in energetics relation between the two systems during the flare impulsive phase.

\begin{figure} 
     \vspace{-0.0\textwidth}    
     \centerline{\hspace*{0.0\textwidth}
               \includegraphics[width=0.77\textwidth,clip=1]{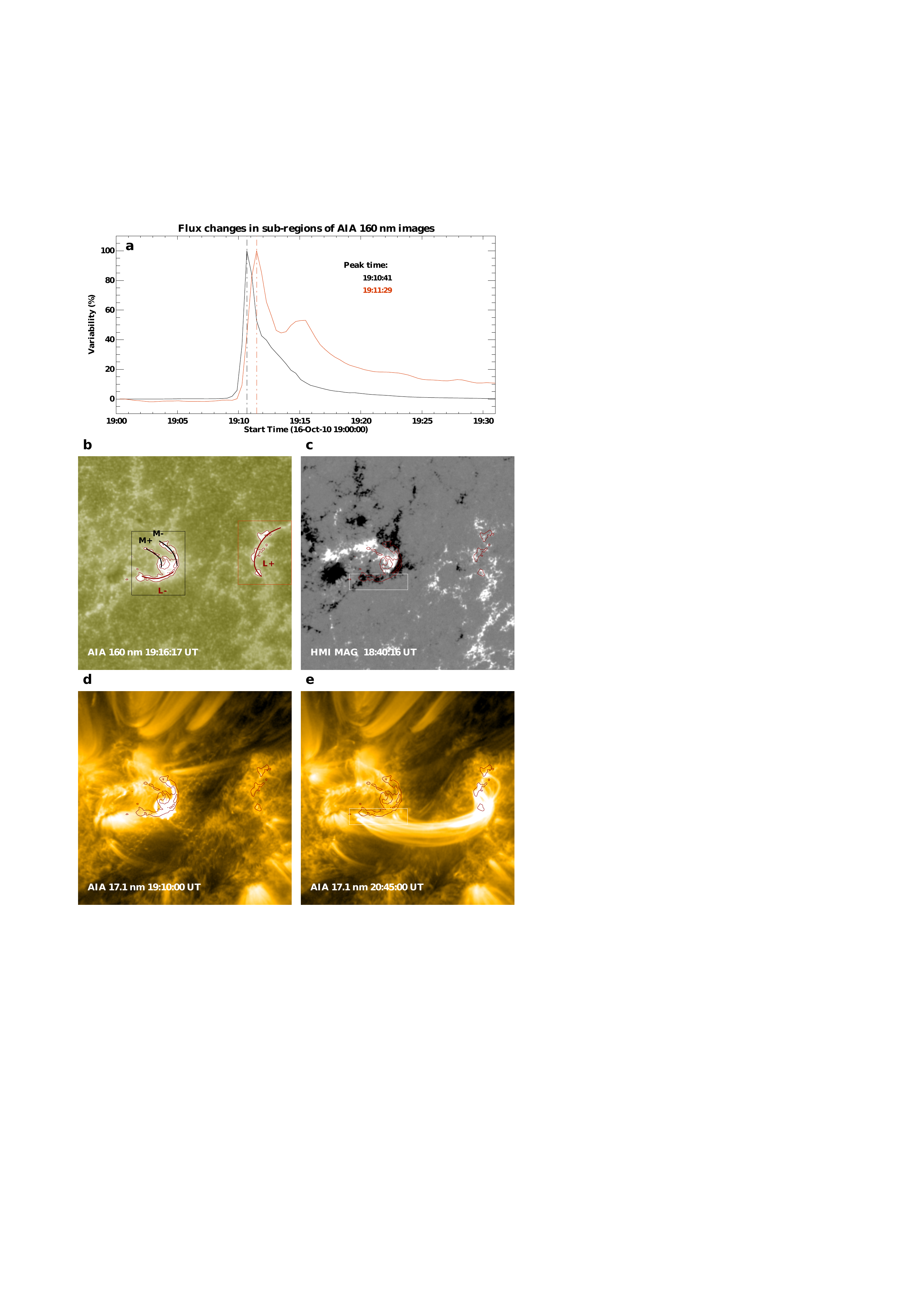}
               }

     \vspace{0.0\textwidth}   

\caption{(a): Intensity profiles of footpoint UV emission from the main region (black) and the late phase region (red) for the M2.9 flare on 2010 October 16. The two regions are indicated by the black and red boxes in panel b respectively; (b): 160 nm image with its own intensity contours outlining the flare ribbons; (c): HMI magnetogram of the active region with AIA 160 nm intensity contours overlaid. (d)-(e): AIA 17.1 nm image with AIA 160 nm intensity contours overlaid.}
   \label{f5}
   \end{figure}

To study the possible relations between the two loop systems, we further inspect the AIA 160 nm transition region images (as shown in Figure~\ref{f5}b) and the HMI magnetogram images (Figure~\ref{f5}c). The transition region images, while not displaying coronal loop structures, better show the brightening footpoints of loop arcades during solar flares. The time sequence of the AIA 160 nm cut-out images (online movie) show largely four brightening ribbons, three of which are located in the flare's main region (inside the black box in Figure~\ref{f5}b), and the fourth one is located in a remote region, corresponding to the isolated footpoint of the late phase arcade (inside the red box in Figure~\ref{f5}b). The three ribbons in the main region are very close in distance and are much brighter than the other one. The two spatially close ribbons (black lines M+ and M- in Figure~\ref{f5}b) are the footpoints of the main phase arcade, and the third ribbon nearby (red line L- in Figure~\ref{f5}b) is one of the footpoints of the late phase arcade. The approximation of these three footpoints makes it difficult to entangle them, and would have appeared as a single elongated ribbon if not because of the high quality AIA data (Figure~\ref{f5}b).

We sum up the total fluxes in the two boxes of the AIA 160 nm images (Figure~\ref{f5}b) and plot their time profiles (Figure~\ref{f5}a). It is interesting to note that the footpoint brightening peak time in the remote region is delayed by about 48 seconds (the uncertainty is 24 seconds, which is the cadence of the measurement for AIA 160 nm passband) from that of the impulsive phase region. This result is intriguing. It may suggest that the energy is injected at the footpoint close to the main phase region, and then transported to the remote footpoint; the delay time may indicate a characteristic transport time. Dividing the delay time from the separating distance of the two footpoints, one can estimate that the transport velocity is $\sim1750$ km/s, which is comparable to the Alfv\'en speed in the corona ($V_A \simeq 1 - 3 \times 10^3$ km/s; \citet{Shibata1992}). This result suggests that the brightening of the remote footpoint of the late phase arcade has a relation with the magnetic disturbance. This disturbance is originated from the compact flare region, where is the site of the magnetic reconnection that produces the impulsive phase energy release.

The two loop arcades seem to be magnetically connected through one of their footpoints from the observations. In Figure~\ref{f5} panel d and e, we show the late phase arcade in AIA 17.1 nm overlaid with the contours of the 160 nm footpoint ribbons, while in Figure~\ref{f5}c the magnetogram image is shown overlaid with the same ribbons. From the magnetogram, one can find that a compact bipolar magnetic field region ($\sim$1000 Gauss, $\sim$200 arcsec${^2}$) with a strong gradient sigmoid polarity inversion line is the surface source of the flare impulsive phase. The two ribbons of the impulsive phase are along each side of the polarity inversion line, where we believe that magnetic reconnection occurred and the flare was produced. On the other hand, the long and high late phase loop arcade has one footpoint (indicated by the white box in Figure~\ref{f5}c and \ref{f5}d) rooted onto a weaker negative polarity region ($\sim$600 Gauss), which is close to the positive magnetic field of the main region. The other footpoint (indicated by the isolated ribbon) is connected to a weaker ($\sim$700 Gauss) positive magnetic polarity region to the west (likely from a decayed active region), which is far away from the main region. These facts suggest that the main phase arcade and the late phase arcade are magnetically connected, even though they are of two different magnetic loop systems (more details will be given in a cartoon model in the Discussion Section).

\subsection{M1.4 flare on 2011 February 18}

\begin{figure} 
     \vspace{-0.0\textwidth}    
     \centerline{\hspace*{0.0\textwidth}
               \includegraphics[width=1.0\textwidth,clip=1]{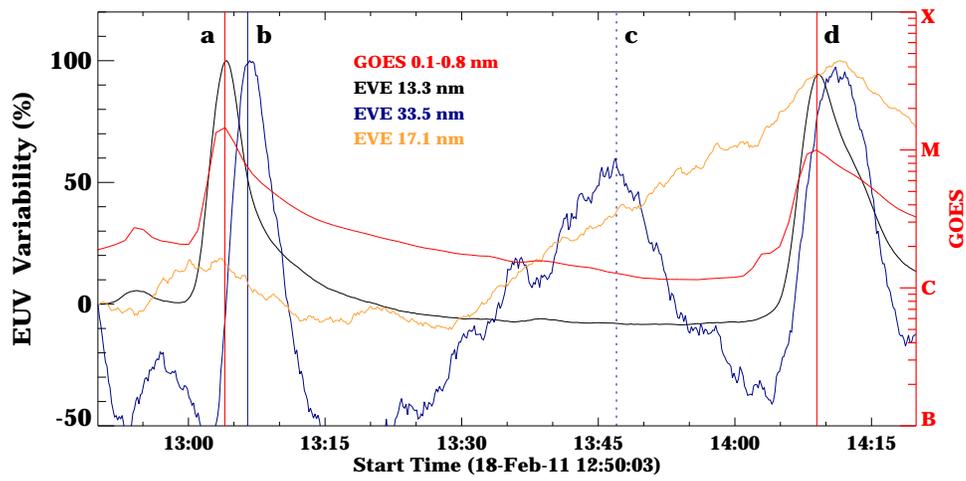}
               }

     \vspace{0.0\textwidth}   

\caption{The temporal variation profiles of the irradiance in three different spectral lines of SDO EVE observation of the M1.4 flare on 2011 February 18. The GOES profile is also shown. Vertical lines a, b, c and d indicates the M1.4 flare's impulsive phase peak, EVE 33.5 nm main phase peak, EVE 33.5 nm late phase peak and the M1.0 flare's impulsive phase peak, respectively.}
   \label{f6}
   \end{figure}

The second event of interest is an M1.4 flare on 2011 February 18. The GOES soft X-ray profile of this flare (red line in Figure~\ref{f6}) shows that its rise phase began at $\sim$13:00 UT and ended $\sim$13:04 UT (vertical solid line a), lasting only for 4 minutes. The gradual phase lasted for about 60 minutes before being contaminated by another independent flare from a different region. This event is also a typical late phase flare since there is an obvious second peak ($\sim$13:47 UT; vertical dotted line c) shown on the EVE 33.5 nm temporal profile (blue line in Figure~\ref{f6}), which is about 41 minutes from its main phase peak ($\sim$13:06 UT; vertical solid line b). Another M1.0 flare, beginning one hour after the onset time of the flare of study, caused the rise of emissions in EVE lines, thus contaminating the late phase light curves of cool temperature lines. Nevertheless, we can use AIA images to remove the contamination in the late phase study.

\begin{figure} 
     \vspace{-0.0\textwidth}    
     \centerline{\hspace*{0.0\textwidth}
               \includegraphics[width=1.0\textwidth,clip=1]{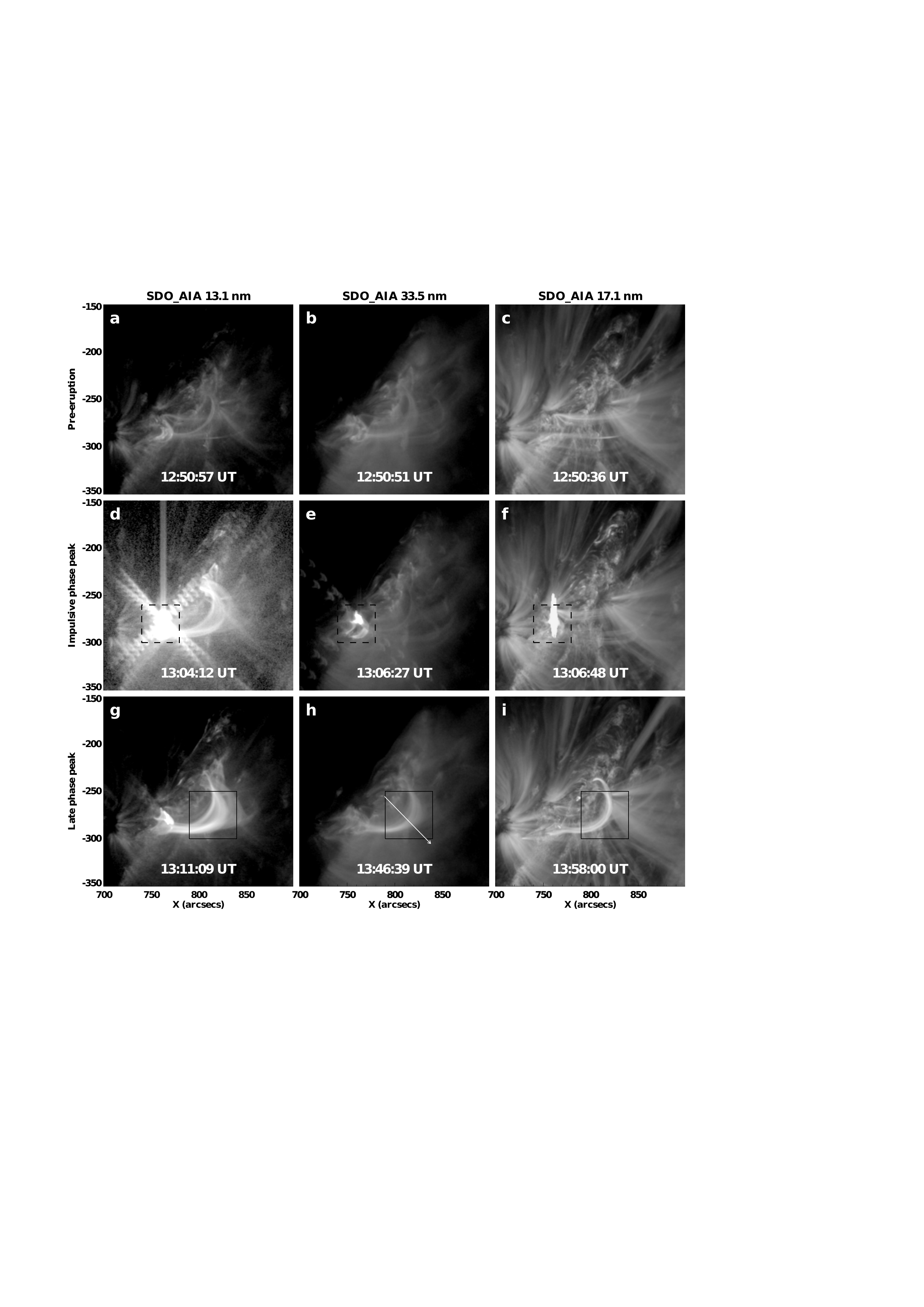}
               }

     \vspace{0.0\textwidth}   

\caption{AIA sub-area images showing the structural evolution of the M1.4 flare on 2011 February 18. From left to right are AIA images in 13.1 nm, 33.5 nm and 17.1 nm passbands, respectively. (a)--(c): the pre-eruption images. (d)--(f): the images at the flare's main phase peak time. The black dotted-line boxes indicate the flare main region. (g)--(i): image at the peak times of the late phase (note that the times are different for different temperatures). The black solid-line boxes indicate the center region of the late phase arcade.}
   \label{f7}
   \end{figure}

\begin{figure} 
     \vspace{-0.0\textwidth}    
     \centerline{\hspace*{0.0\textwidth}
               \includegraphics[width=0.77\textwidth,clip=1]{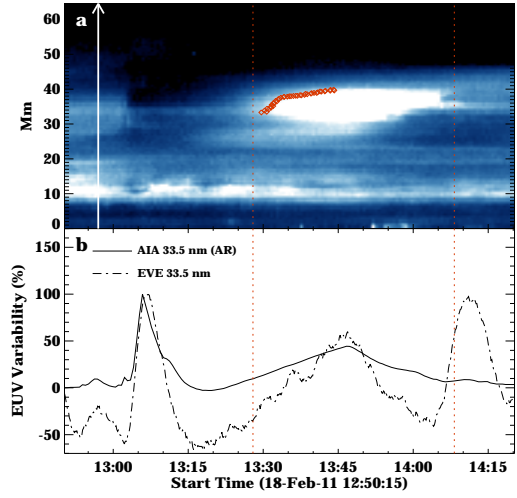}
               }

     \vspace{0.0\textwidth}   

\caption{(a): Slice-time sequences of the AIA 33.5 nm images for the flare on 2011 February 18. The slice is indicated by the arrow line in Figure ~\ref{f7}h. (b): EUV variability from both EVE 33.5 nm and AIA 33.5 nm observations. The AIA flux is the total flux of the cut-out of the AR shown in Figure ~\ref{f6}.}
   \label{f8}
   \end{figure}

From the AIA images as shown in Figure~\ref{f7}, we find that the M1.4 flare was originated from NOAA AR 11158, centered at 20${^\circ}$S and 53$^{\circ}$E when the flare occurred. The other contaminating M1.0 flare was originated from NOAA AR 11162, at 18 N${^\circ}$ and 02 W$^{\circ}$. Thus the two flares are spatially separated. We cut out a small region that contains the whole AR 11158 from the AIA images as we did for the first event. In the images, there also appear two sets of loop arcades, corresponding to the flare main phase and the late phase, respectively. The main phase arcade (inside the black dotted-line box in Figure~\ref{f7}) has a linear span of $\sim$30 arcsec, or 21,500 km between its two footpoints. On the other hand, the late phase arcade (indicated by the black solid-line box in Figure~\ref{f7}) has a span of $\sim$100 arcsec, or 71,700 km between the two footpoints. The late phase arcade is more than three times as large as the main phase arcade. Further, the late phase arcade lies above the main phase arcade. The late phase arcade appears at the same location in all temperatures, and has an expansion along the slice direction as shown in the slice-time plot of Figure~\ref{f8}. This Figure also shows again that the EUV late phase originates from the large loop arcade seen in AIA images.

\begin{figure} 
     \vspace{-0.0\textwidth}    
     \centerline{\hspace*{0.0\textwidth}
               \includegraphics[width=1.0\textwidth,clip=1]{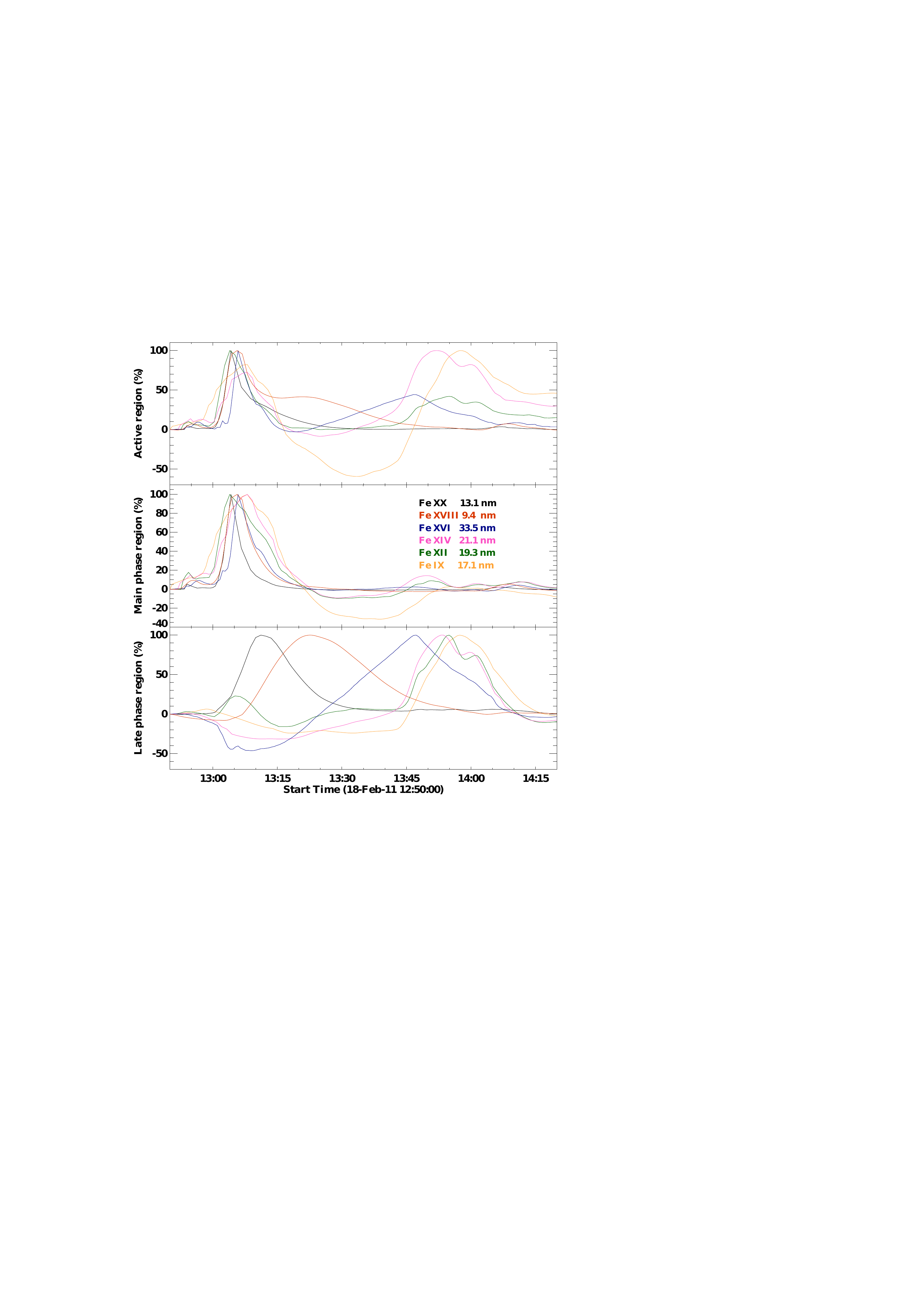}
               }

     \vspace{0.0\textwidth}   

\caption{Intensity profiles of total EUV flux in selected regions of M1.4 flare on 2011 February 18 based on AIA images of six coronal passbands. The top panel shows the profiles of the whole active region. The middle and bottom panels show that from the flare main phase region and the flare late phase region, respectively.}
   \label{f9}
   \end{figure}

Similar to the first event, we plot the EUV light curves of three different regions in Figure~\ref{f9} and list the measured values in Table~\ref{t2}. Figure~\ref{f9} shows that the thermal evolution of the flare late phase is very similar to that of the first event. While the main phase brightenings in all temperatures peaked within several minutes of each other, the peak times of the late phase are much different in different temperatures. The time delays between the late phase peaks of the six passbands with respect to the impulsive phase peak are 7, 19, 42, 51, 52, and 54 minutes respectively from high to low temperatures.

\begin{figure} 
     \vspace{-0.0\textwidth}    
     \centerline{\hspace*{0.0\textwidth}
               \includegraphics[width=0.77\textwidth,clip=1]{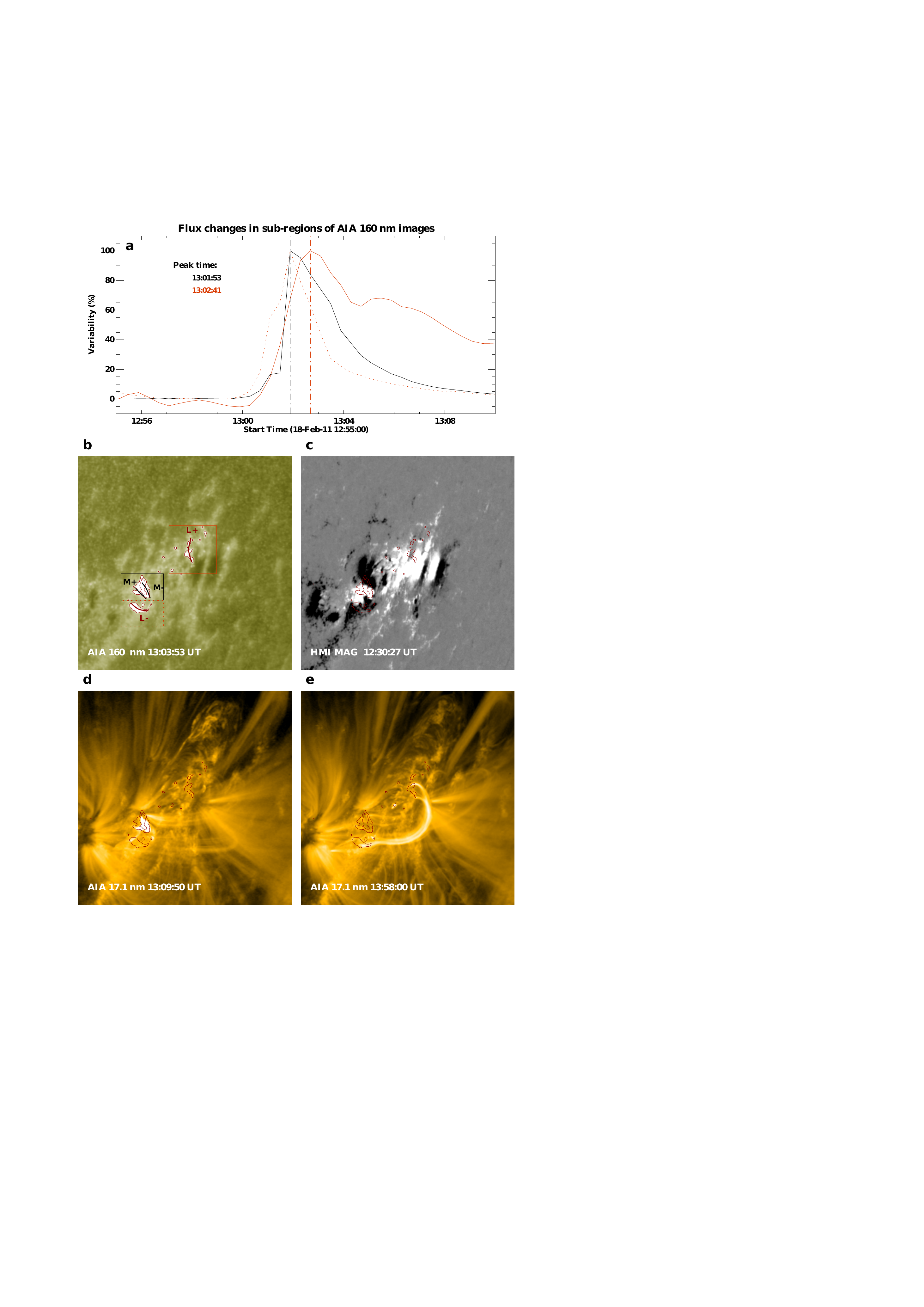}
               }

     \vspace{0.0\textwidth}   

\caption{(a): Intensity profiles of footpoint UV emission for the main region (black) and the late phase region (red) for the M1.4 flare on 2011 February 18. The two regions are indicated by the black and red boxes in panel b respectively; (b): 160 nm image with its own intensity contours outlining the brightening ribbons; (c): HMI magnetogram of the active region with AIA 160 nm intensity contours overlaid. (d)-(e): AIA 17.1 nm image with AIA 160 nm intensity contours overlaid.}
   \label{f10}
   \end{figure}

There are clearly four footpoint ribbons appearing in the AIA 160 nm images (Figure~\ref{f10} and the online movie). The two pairs of ribbons show better in this event than the previous event. The two impulsive phase ribbons (black lines M+ and M- in Figure~\ref{f10}b) were much brighter in their appearance. The third ribbon (red lines L- in Figure~\ref{f10}b), which was much closer to the impulsive phase region than the fourth ribbon (red line L+ in Figure~\ref{f10}b), appeared at the same time as the impulsive phase ribbons (black dot dash vertical line in Figure~\ref{f10}a). The fourth ribbon was located at the isolated footpoint of the late phase loop arcade and the peak of its brightness is delayed about 48 seconds with respect to the peak time of the other ribbons (red dot dash vertical line in Figure~\ref{f10}a); this delay time denotes a transport speed of $\sim1500$ km/s. This result, similar to that of the first event, suggests that both loop arcades are heated by a same energy source. Therefore, the flare late phase has a physical connection with the impulsive phase. The magnetic topology of this AR is complicated, including several polarity inversion lines (Figure~\ref{f10}c). Combining the observations of AIA and HMI, we find that the impulsive phase arcade straddles above the sigmoid polarity inversion line in between the two impulsive phase ribbons. Thus, the flare is produced in the strong magnetic region ($\sim$1200 Gauss) associated with this polarity inversion line. The lower footpoint of the late phase arcade is rooted at the weaker negative magnetic region ($\sim$800 Gauss), which is near the positive magnetic region of the impulsive phase arcade, while the other footpoint, the isolated one, is rooted at a weaker positive magnetic field ($\sim$700 Gauss) far away from the impulsive phase region.

\section{Discussions}

\begin{figure} 
     \vspace{-0.0\textwidth}    
     \centerline{\hspace*{0.0\textwidth}
               \includegraphics[width=1.0\textwidth,clip=1]{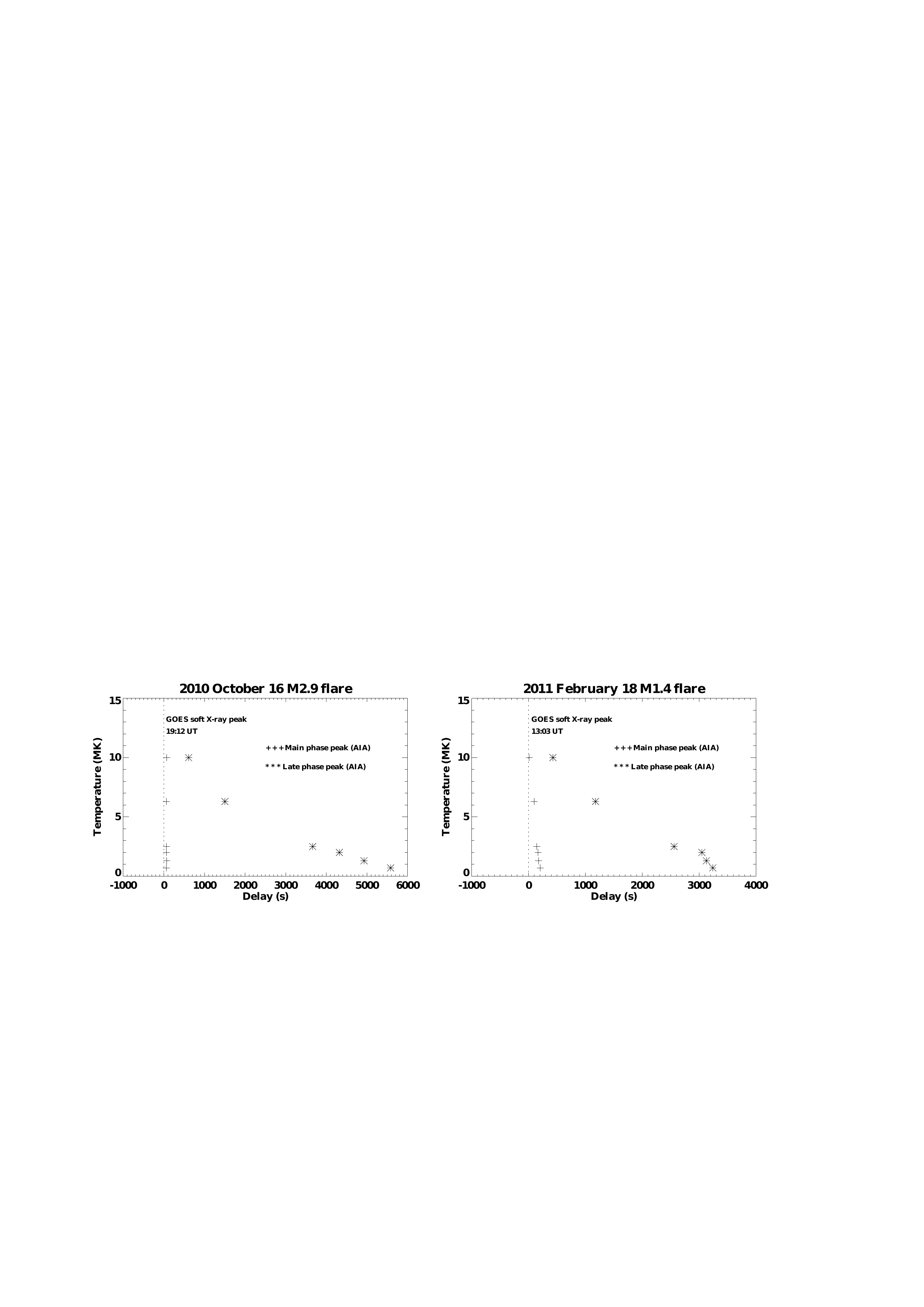}
               }

     \vspace{0.0\textwidth}   

\caption{Temperature-delay plots showing that the delay times of the EUV flux peak in different temperatures with respect to the soft X-ray flux peak. The cross symbols show the delay of the flare main phase, while the asterisk symbols for the flare late phase.}
   \label{f11}
   \end{figure}

The fact that the late phase arcade appears progressively later in time from high to low temperatures indicates a prolonged cooling process. We plot in Figure ~\ref{f11} the delays between the GOES soft X-ray peak time and the EUV peak times in different temperature for both the main phase and the late phase for the two events studied in the paper. It shows that the late phase has a much longer cooling delay from hot to cool temperatures than that in the main phase. The delay times of the late phase peaks in Fe IX 17.1 nm (0.7 MK) for the two events are about 5700 s and 3300 s respectively. These values are much longer than the typical radiation/conduction cooling time (a few hundred seconds; \citet{Svestka1987}) of flares. Nevertheless, we can estimate the cooling time of the late phase loop arcade by using the formula derived by \citet{Cargill1995} based on a simple cooling model.

 $$\tau _{cool}\doteq 2.35\times 10^{-2}L^{5/6}T_{e}^{-1/6}n_{e}^{-1/6}$$

\noindent which assumes that during the evolution of a flaring loop, the conductive cooling dominates initially and then radiative cooling takes over. The parameters L, $T_{e}$ and $n_{e}$ are, respectively, the loop half-length, electron temperature, and electron density at the beginning of the cooling.

From the temporal profiles of the fluxes of the AIA 160 nm images (Figure~\ref{f5}a,~\ref{f10}a), we infer that the late phase loop arcades should be heated at the same time as the main phase loops for the footpoints at the flare main region, and should be within one minute for the other footpoint. This indicates that the initial plasma temperature of the late phase arcade must be higher than $1.0\times10^{7} K$, since the peak brightness of $1.0\times10^{7} K$ plasma (Fe XX 13.1 nm) is delayed by about 9 minutes from the soft X-ray peak. We assume that the initial plasma temperature of approximately $2.5\times10^{7} K$, which is about the temperature of the soft X-ray emission \citep{Thomas1985}. The density can be determined from the total emission measure (EM, defined as $\int n_{e}^{2}dL$ along the line of sight, where L is the length of the emission plasma along the line of sight). Using all six coronal passbands of AIA images, we measure the EM of the late phase loop arcade and derive the electron density; it is found that the density is $\sim2-10\times10^{9}$ cm$^{-3}$ for the first event and $\sim4-10\times10^{9}$ cm$^{-3}$. Note that a detailed discussion of the EM method is beyond the scope of this paper, and one can refer to the recent studies of \citet{Schmelz2011} and \citet{Cheng2012} for detail. In any case, our density estimation is semi-quantitative, which is relatively insensitive to the EM models used. The loop half-length is determined from the span of the two footpoints of the loop arcade, which gives $7\times10^{9}$ cm and $3.6\times10^{9}$ cm for the two events respectively. These parameters yield a cooling time of $\sim4700-6200$ s for the first event and $\sim2700-3200$ s for the second event, which is qualitatively consistent with the observed delay times. Therefore, we may argue that the time of energy injection into the late phase arcade is very close to that of the main phase. The late appearance of the late phase loop arcade in EUV is mainly a cooling-delay effect, rather than the effect of a later energy injection.

We further argue that there might be a cause-effect relationship between the flare main phase and the late phase, even though the two phases originate from two different magnetic loop systems. From the AIA 160 nm images, we found that the flare main phase loop arcade footpoints brighten at almost the same time as the nearby late phase loop arcade footpoint, suggesting that they are energized at almost the same time. It has been well accepted that the flare energy is produced by the fast magnetic reconnection occurring in the corona \citep{Priest2002}. Similarly, we can assume that the late phase flare energy is also produced by magnetic reconnection. If not sharing one common reconnection, the two reconnections should be related with each other and very close in time. This argument is reinforced by the fact that the two magnetic loop systems are connected in topology.

\begin{figure} 
     \vspace{-0.0\textwidth}    
     \centerline{\hspace*{0.0\textwidth}
               \includegraphics[width=1.0\textwidth,clip=1]{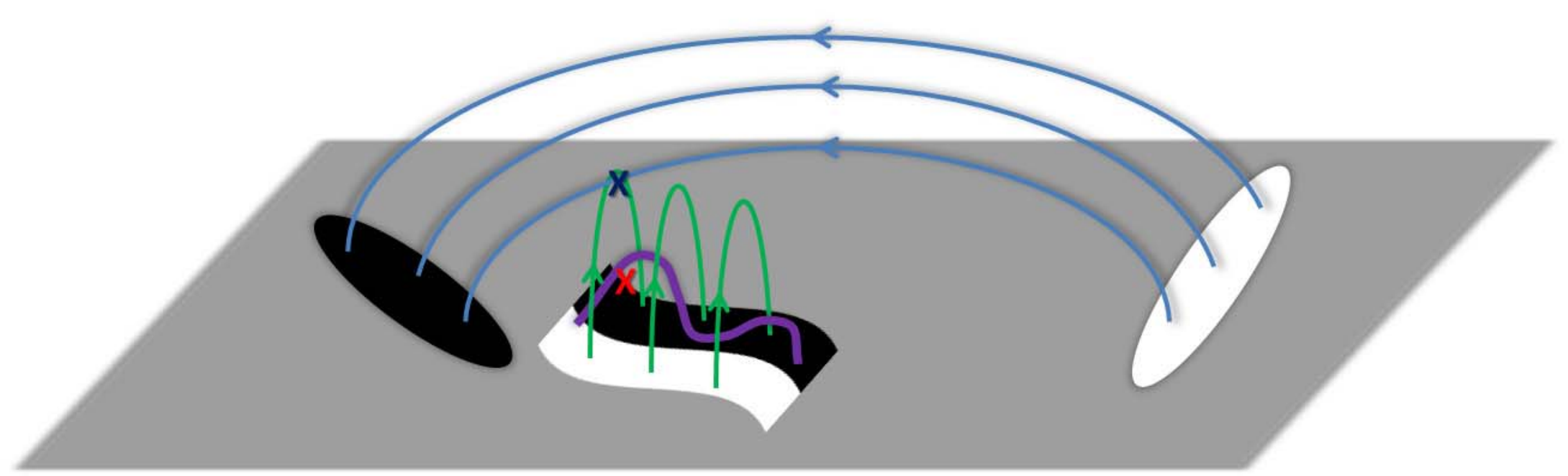}
               }

     \vspace{0.0\textwidth}   

\caption{A schematic of the magnetic topology of an EUV late phase flare. The purple twisted lines indicate the sheared core magnetic field stretching over the polarity inverse line of the compact dipole; the red `X' indicates the site of flare reconnections underneath the rising core, while the black `X' indicates the site of the second magnetic reconnection between the small active region arcade (green) over the sheared core with the large loop arcade (blue). }
   \label{f12}
   \end{figure}

Such topological connection can be illustrated in a simple conceptual cartoon model as shown in Figure~\ref{f12}. There is an asymmetric quadruple magnetic flux system in this model. The compact dipole which contains a sigmoidal core produces the main phase flare through a process like in the standard flare model. Magnetic reconnection happens underneath the pre-existing sheared core field lines (not excluding the possibility that there exists a magnetic flux rope in the core; the red `X' indicates the location of the flare reconnection; the twisted purple line indicates the sheared core field). As the sheared core field/flux rope rises and erupts, the magnetic reconnection underneath forms the post-flare loop arcade, i.e. the main phase loop arcade. Given the quadruple magnetic topology in the model, the rising active region field will inevitably interact and reconnect with the overlying large loop arcade, leading to the second magnetic reconnection as shown by the black `X' symbol). Due to the fact that the flare main phase region is close to one footpoint of the large loop arcade, the two reconnections appear almost simultaneously. After the reconnections, the small arcade cools down quickly, while the large arcade cools down more slowly, which produces the observed late phase.

\citet{Woods2011} also pointed out that the EUV late phase may be caused by another reconnection, which has a different location and rate from the flare reconnection. In this study, we further propose that there is a cause-effect relationship between the two reconnections. From the observations of Large Angle and Spectrometric Coronagraph (LASCO) onboard of Solar and Heliospheric Observatory (\emph{SOHO}), one can find that the M2.9 flare is eruptive (associated with a CME) and the M1.0 flare is not associated with any CME. However, both flares display the EUV late phase and are very similar in post-eruption topology. These facts suggest that the existence of the late phase does not depend on whether or not the flare is eruptive but rather on the magnetic topology, i.e., as long as the second magnetic reconnection occurs there would be a late phase.

Note that the late phase arcade can be clearly seen in hot temperature AIA images (e.g. 13.1 nm, Figure~\ref{f2}h and \ref{f6}h). However, the late phase is not obvious in the profiles of EVE hot lines (e.g. 13.3 nm, Figure~\ref{f1} and \ref{f6}), although it is conspicuous in warm lines. The reason for the absence of the late phase in EVE hot lines is that the EUV irradiances of the late phase arcade in hot lines is relatively small compared to that of the impulsive phase, i.e., 4\% for the first event (Table~\ref{t1}) and 14\% for the second event as derived from the AIA 13.1 nm (Table~\ref{t2}) imaging observations. The late phase is further obscured by the closeness in time of the two peaks in hot temperatures; the EUV emission is still dominated by that of the impulsive phase when the emission in the late phase arcade reaches the peak brightness. On the other hand, the ratios of the brightness between the late phase peak and the impulsive peak are 15$\%$ and 36$\%$ for the two events respectively. In cooler lines, the EUV irradiances are dominated by the bulk corona emission instead of flare emission, making the late phase emission less obvious. Therefore, the warm temperature of EVE 33.5 nm line is the best for detecting the presence of the late phase of a flare.

\section{Conclusion}

Based on the study of the two late phase flares, an M2.9 flare on 2010 October 16 and an M1.4 flare on 2011 February 18, we find that the flare's EUV late phase originates from a different loop arcade from the main phase loop arcade in an active region. Nevertheless, from the topological point of view, the two loop arcades, one extended and the other compact in size, are magnetically connected. One footpoint of the late phase arcade adjoins the magnetic region of the compact main phase arcade, forming an asymmetric quadruple magnetic configuration. We further find that the late phase arcade appears in hot temperatures first, minutes after the impulsive phase peak, then progressively cools down to lower and lower temperatures; this cooling process lasts for more than one hour from the hot temperature of $\sim10 MK$ to the cool temperature of $\sim1 MK$. On the other hand, the delay times from hot to cool temperatures in the main phase are significantly shorter, e.g., in the order of a few minutes.
We argue that the late phase originates from a second magnetic reconnection between the rising main flare magnetic field and the large loop arcade. The second magnetic reconnection immediately follows the onset of the flare reconnection.

Finally, it is worth mentioning that the scenario of the EUV late phase described above poses a challenge to the standard flare model ~\citep{Svestka1992}. In the standard model, only one bipolar system is involved, with a flux rope and/or a highly sheared magnetic field at its core region. The flare produces a pair of separating ribbons and a rising loop arcade. On the other hand, a late phase flare must involves two bipolar systems, since it produces two loop arcades in the corona and two pairs of ribbons in the lower atmosphere. Thus, for better understanding the flare process that involves the late phase, it is necessary to modify the standard flare model to incorporate two bipolar magnetic loop systems and model the energy release process through magnetic reconnections that possibly involves both systems.


\acknowledgements
SDO is a mission of NASA's Living With a Star Program. K.L and Y. W. are supported by 973 key project 2011CB811403 and NSFC 41131065. K.L. is also supported by the scholarship granted by the China Scholarship Council (CSC) under file No. 2011634065. J.Z., is supported by NSF grant ATM-0748003, AGS-1156120 and NASA grant NNG05GG19G.


\bibliographystyle{apj}

\begin{thebibliography}{}
\expandafter\ifx\csname
natexlab\endcsname\relax\def\natexlab#1{#1}\fi



\bibitem[Battaglia et al.(2008)]{Battaglia2008} Battaglia, M., Fletcher, L., \& Benz, A.~O.\ 2008, 12th European Solar Physics Meeting, Freiburg, Germany, held September, 8-12, 2008, p.2.85, 12, 2

\bibitem[Benz(2008)]{Benz2008} Benz, A.~O.\ 2008, Living Reviews
in Solar Physics, 5, 1

\bibitem[Cargill et al.(1995)]{Cargill1995} Cargill, P.~J.,
Mariska, J.~T., \& Antiochos, S.~K.\ 1995, \apj, 439, 1034

\bibitem[Carmichael(1964)]{Carmichael1964} Carmichael, H.\ 1964, NASA
Special Publication, 50, 451

\bibitem[Carrington(1859)]{Carrington1859} Carrington, R.~C.\ 1859,
\mnras, 20, 13

\bibitem[Cheng et al.(2012)]{Cheng2012} Cheng, X., Zhang, J.,
Saar, S.~H., \& Ding, M.~D.\ 2012, \apj, 761, 62

\bibitem[Fletcher et al.(2011)]{Fletcher2011} Fletcher, L., Dennis,
B.~R., Hudson, H.~S., et al.\ 2011, \ssr, 159, 19

\bibitem[Hirayama(1974)]{Hirayama1974} Hirayama, T.\ 1974, \solphys,
34, 323

\bibitem[Hudson(2011)]{Hudson2011} Hudson, H.~S.\ 2011, \ssr, 158,
5

\bibitem[Kane et al.(1980)]{Kane1980} Kane, S.~R., Crannell,
C.~J., Datlowe, D., et al.\ 1980, Skylab Solar Workshop II, 187

\bibitem[Kopp
\& Pneuman(1976)]{Kopp1976} Kopp, R.~A., \& Pneuman, G.~W.\ 1976, \solphys, 50, 85

\bibitem[Lemen et al.(2011)]{Lemen2011} Lemen, J.~R., et al.\
2011, \solphys, 106

\bibitem[Moore et al.(1980)]{Moore1980} Moore, R., McKenzie,
D.~L., Svestka, Z., et al.\ 1980, Skylab Solar Workshop II, 341

\bibitem[Neupert(1968)]{Neupert1968} Neupert, W.~M.\ 1968, \apjl,
153, L59

\bibitem[Pesnell et al.(2012)]{Pesnell2012} Pesnell, W.~D.,
Thompson, B.~J., \& Chamberlin, P.~C.\ 2012, \solphys, 275, 3

\bibitem[Priest
\& Forbes(2002)]{Priest2002} Priest, E.~R., \& Forbes, T.~G.\ 2002, \aapr, 10, 313

\bibitem[Scherrer et al.(2012)]{Scherrer2012} Scherrer, P.~H.,
Schou, J., Bush, R.~I., et al.\ 2012, \solphys, 275, 207

\bibitem[Schmelz et al.(2011)]{Schmelz2011} Schmelz, J.~T., Worley,
B.~T., Anderson, D.~J., et al.\ 2011, \apj, 739, 33

\bibitem[Shibata et al.(1992)]{Shibata1992} Shibata, K., Ishido,
Y., Acton, L.~W., et al.\ 1992, \pasj, 44, L173

\bibitem[Sturrock(1968)]{Sturrock1968} Sturroc P.A., A model of solar flares. Kiepenheur (ed.), Structure and development of solar active regions. IAU Symposium no. 35, 471 (1968)

\bibitem[Svestka(1987)]{Svestka1987} Svestka, Z.\ 1987, \solphys,
108, 411

\bibitem[Svestka
\& Cliver(1992)]{Svestka1992} Svestka, Z., \& Cliver, E.~W.\ 1992, IAU Colloq.~133: Eruptive Solar Flares, 399, 1

\bibitem[Thomas et al.(1985)]{Thomas1985} Thomas, R.~J., Crannell,
C.~J., \& Starr, R.\ 1985, \solphys, 95, 323

\bibitem[Woods et al.(2011)]{Woods2011} Woods, T.~N., Hock, R.,
Eparvier, F., et al.\ 2011, \apj, 739, 59

\bibitem[Woods et al.(2012)]{Woods2012} Woods, T.~N., Eparvier,
F.~G., Hock, R., et al.\ 2012, \solphys, 275, 115



\end{thebibliography}




\begin{table}
\tiny
\setlength{\tabcolsep}{4pt}
\caption{Properties of the EUV light curves of 2010 October 16 M2.9 flare}
\begin{center}
\begin{tabular*}{1.0\textwidth}{l l c c c c c c c c}
\hline\hline
Passband & Ion & Temperature & \multicolumn{2}{c}{Peak time (UT)}  & \multicolumn{2}{c}{Delay time (min)} & \multicolumn{2}{c}{FWHM (min)}  & Ratio \\
(nm) &  & (MK) & Main Phase & Late Phase & Main Phase & Late Phase  & Main Phase & Late Phase & P$_{LP}$/P$_{MP}$ \\
\hline
13.1\footnotemark[1] & Fe XX & 10 & 19:13:13 & 19:22:09 & 0 & 9  & 4 & 22 & 0.04 \\
9.4 & Fe XVIII & 6.3 & 19:13:02 & 19:37:02 & 0 & 24  & 5 & 32 & 0.07 \\
33.5 & Fe XVI & 2.5 & 19:13:03 & 20:13:03 & 0 & 60 & 6 & 55 & 0.15 \\
21.1 & Fe XIV & 2.0 & 19:13:03 & 20:24:00 & 0 & 71 & 6 & 37 & 0.25 \\
19.3 & Fe XII & 1.3 & 19:13:10 & 20:34:07 & 0 & 81 & 4 & 33 & 0.12 \\
17.1 & Fe IX & 0.7 & 19:13:00 & 20:45:00 & 0 & 92 & 7 & 34 & 0.28 \\
\hline
\end{tabular*}
\end{center}
\label{t1}
\end{table}
\footnotetext[1]{AIA 13.1 nm passband detects emissions from both hot ($\sim$10 MK) and cool ($\sim$0.5 MK) plasmas. In this study, the enhanced emission should be from the hot plasma of the flare, which is consistent with the EVE 13.3 nm observation.}


\begin{table}
\tiny
\setlength{\tabcolsep}{4pt}
\caption{Properties of the EUV light curves of 2011 February 18 M1.4 flare}
\begin{center}
\begin{tabular*}{1.0\textwidth}{l l c c c c c c c c}
\hline\hline
Passband & Ion & Temperature & \multicolumn{2}{c}{Peak time (UT)}  & \multicolumn{2}{c}{Delay time (min)}  & \multicolumn{2}{c}{FWHM (min)}  & Ratio \\
(nm) &  & (MK) & Main Phase & Late Phase & Main Phase & Late Phase & Main Phase & Late Phase & P$_{LP}$/P$_{MP}$ \\
\hline
13.1 & Fe XX& 10 & 13:04:12 & 13:11:09 & 0 & 7  & 3 & 12 & 0.14 \\
9.4 & Fe XVIII & 6.3 & 13:05:38 & 13:23:38 & 2 & 19  & 4 & 23 & 0.25 \\
33.5 & Fe XVI & 2.5 & 13:06:27 & 13:46:39 & 3 & 42 & 4 & 38 & 0.36 \\
21.1 & Fe XIV & 2.0 & 13:06:48 & 13:54:48 & 3 & 51 & 6 & 20 & 0.59 \\
19.3 & Fe XII & 1.3 & 13:06:57 & 13:56:07 & 3 & 52 & 5 & 24 & 0.53 \\
17.1 & Fe IX & 0.7 & 13:07:25 & 13:58:00 & 4 & 54 & 8 & 17 & 0.65 \\
\hline
\end{tabular*}
\end{center}
\label{t2}
\end{table}


\end{document}